\documentclass[conference]{IEEEtran}
\usepackage{epsfig,endnotes}
\usepackage{booktabs} 
\usepackage{caption}
\usepackage{cleveref}
\usepackage[breaklinks,colorlinks]{hyperref}
\hypersetup{citecolor=blue,linkcolor=blue,urlcolor=black}
\usepackage{subcaption}

\usepackage{booktabs} 

\usepackage[procnames]{listings}
\usepackage{xcolor}

\definecolor{mygreen}{RGB}{0,204,153}
\definecolor{myblue}{RGB}{0,0,255}
\definecolor{mygray}{rgb}{0.5,0.5,0.5}
\definecolor{mymauve}{rgb}{0.58,0,0.82}
\definecolor{mystring}{RGB}{66, 134, 244}
\definecolor{mydiffminus}{RGB}{239, 66, 43}

\lstdefinestyle{custompy}{
  belowcaptionskip=1\baselineskip,
  breaklines=true,
  frame=single,
  xleftmargin=\parindent,
  showstringspaces=false,
  basicstyle=\footnotesize\ttfamily,
  keywordstyle=\bfseries\color{myblue},
  commentstyle=\itshape\color{mymauve},
  stringstyle=\color{mystring},
  procnamekeys={def},
  procnamestyle=\color{myblue},
  upquote=true,
}

\lstdefinelanguage{diff}{
  morecomment=[f][\color{magenta}]-,         
  morecomment=[f][\color{mystring}]+,       
  morecomment=[f][\color{magenta}]{---}, 
  morecomment=[f][\color{mystring}]{+++},
}

\begin{document}

\title{Rethinking Misalignment to Raise the Bar for Heap Pointer Corruption}

\author{\IEEEauthorblockN{Daehee Jang}
\IEEEauthorblockA{KAIST\\
daehee87.kr@gmail.com}
\and
\IEEEauthorblockN{Hojoon Lee}
\IEEEauthorblockA{KAIST\\
hjlee229@kaist.ac.kr}
\and
\IEEEauthorblockN{Brent Byunghoon Kang}
\IEEEauthorblockA{KAIST\\
brentkang@kaist.ac.kr}}

\maketitle

\begin{abstract}
Heap layout randomization renders a good portion of heap vulnerabilities unexploitable. However, some remnants of the vulnerabilities are still exploitable even under the randomized layout. According to our analysis, such heap exploits often abuse pointer-width allocation granularity to spray crafted pointers. To address this problem, we explore the efficacy of byte-granularity (the most fine-grained) heap randomization. Heap randomization, in general, has been a well-trodden area; however, the efficacy of byte-granularity randomization has never been fully explored as \emph{misalignment} raises various concerns. This paper unravels the pros and cons of byte-granularity heap randomization by conducting comprehensive analysis in three folds: (i) security effectiveness, (ii) performance impact, and (iii) compatibility analysis to measure deployment cost. Security discussion based on 20 CVE case studies suggests that byte-granularity heap randomization raises the bar against heap exploits more than we initially expected; as pointer spraying approach is becoming prevalent in modern heap exploits. Afterward, to demystify the skeptical concerns regarding misalignment, we conduct cycle-level microbenchmarks and report that the performance cost is highly concentrated to edge cases depending on L1-cache line. Based on such observations, we design and implement an allocator suited to optimize the performance cost of byte-granularity heap randomization; then evaluate the performance with the memory-intensive benchmark (SPEC2006). Finally, we discuss compatibility issues using Coreutils, Nginx, and ChakraCore.

\end{abstract}

\section{Introduction}
\label{s:intro}
Memory corruption vulnerabilities are widely exploited as attack vectors. According to the recent statistics from the Common Vulnerabilities and Exposures (CVE) database, the majority of the arbitrary code execution exploits that have a CVSS score greater than 9.0 are caused by heap-based vulnerabilities such as use-after-free and heap overflow~\cite{cve}. Recent vulnerability reports~\cite{report} also suggest that most of the usefully exploitable bugs are typically caused by heap corruptions.  So far, numerous heap randomization approaches have been proposed. However, some remnants of vulnerabilities still survive due to their exploitation primitives. In modern heap exploit attacks, type confusions between integer-pointer or float-pointer are often utilized~\cite{cve:2017-5030,cve:2017-2521}. To trigger such confusion, a crafted object/pointer spraying technique is required. These advanced heap exploitation techniques take advantage of the fact that although the overall heap layout is unpredictable, the pointer-width alignment of any heap chunk is deterministic (the chunk address is always divisible by \texttt{sizeof(void*)}).

Several heap-related defense approaches, including~\cite{iyer2010preventing,berger2006diehard,novark2010dieharder,akritidis2010cling,bhatkar2003address,kharbutli2006comprehensively,gorenc2015abusing}, have provided insights into making the heap exploitation more difficult by effectively randomizing the heap layout. However, none of the previous methods considered reducing this randomization granularity into byte-level as it breaks the ``CPU word alignment'' for heap memory access. For example,~\cite{kharbutli2006comprehensively,zhong2011dice} and~\cite{bhatkar2003address} randomizes the heap layout by prepending a random sized dummy memory space between each heap chunk; however, the randomized distance between heap chunk is guaranteed to be divisible by \texttt{sizeof(void*)} to respect the CPU word granularity memory alignment. In the case of~\cite{ding2010heap}, the paper suggests the importance of reducing the memory allocation granularity\footnote{To quote the paper: ``memory managers at all levels should use finer memory allocation granularity for better security''} for heap defense. However, the paper considers the pointer-width granularity (8-bytes) as the smallest possible allocation in their discussion. 

%

Regardless of the heap randomization entropy, unless the allocation granularity is bigger than the width of a pointer, the possibility of such an overlapping event increases as the attacker expands the out-of-bounds accessed\footnote{In this paper, out-of-bounds access implies inter-chunk memory access that crosses the heap chunk boundary, including the use of dangling pointers.} heap region inside the target heap segment. In many cases, the adversary has no restriction as to the size of sprayed data. Byte granularity randomization imposes constant probability of failure for dereferencing of any attacker-crafted pointer inside the out-of-bounds heap area. 

Heap exploitation often accompany a spraying a large chunk of memory with a payload that has crafted pointers in it. These malicious pointers crafted by the attacker will overwrite good pointers inside heap (out-of-bounds write vulnerability) or dereferenced by benign codes posing as intact pointers (out-of-bounds read vulnerability). Due to the spatial locality, the attacker is likely to control the relative layout of her spray and target region. Once the two region overlaps, pointer manipulation reliably works as they share the same word alignment.

To mitigate such attack techniques, we explore the efficacy of byte-granularity heap randomization (eliminating the predictability of any memory alignment) and show its detailed results and numerous findings. The need for byte-granularity heap randomization is also inspired by statements of the security researchers at Google Project-Zero, or Pwn2Own contest winners who often emphasized the accurate prediction of word alignment and chunk allocation granularity for making the exploitation possible and reliable\footnote{``How the original type object members align with the confused type object members has a big impact on the exploitability and the reliability''~\cite{bloga}}. 

The effectiveness of byte granularity heap randomization is based on the observation that the majority of heap exploit attacks inevitably involves \emph{pointer corruption} with attacker-crafted values such as string, integer or float; and confusing the application to dereference such attacker-crafted values as intact pointers. Since the width of a pointer is universally known, and the heap chunk at a random location address always follows multiples of this allocation granularity, the attacker can spray her crafted-value repeatedly into the out-of-bounds heap region without calculating its exact location, and defragment the heap until the attacker's crafted-pointer exactly overlaps the benign target pointer. This effect will be illustrated with a figure in~\autoref{s:main}.

According to our analysis and experiments, the efficacy of byte-granularity heap randomization for exploit mitigation can be stronger than simply hindering a single dereferencing of crafted pointer. The reason is that because advanced heap exploits often require a set of \emph{crafted pointer chains}. To show the security effectiveness of byte-granularity heap randomization, we use public heap corruption vulnerabilities which enabled successful attack demonstration in Pwn2Own contest, bug bounty programs, or real-world malware. We use the corresponding CVEs as case studies and explore how byte-granularity heap randomization would affect the exploitability of such vulnerabilities. 

Aside from security issues, byte-granularity heap randomization can be considered skeptical regarding its practicality due to hardware-level limitation as it involves unaligned memory access. However, major CPU vendors such as Intel and ARM started putting efforts to support arbitrarily misaligned (byte-granularity) memory access from hardware level~\cite{agner,armv6}. Based on instruction-level microbenchmark results, we designed an allocator which adopts byte-granularity randomness to each heap chunk address while minimizing the performance penalty induced by unaligned access. We name our allocator ``Randomly Unaligned Memory Allocator (RUMA)''. Performance regarding chunk management and allocation speed of RUMA does not outperform traditional heap allocators. However, RUMA leverages architecture specifics and reduces memory access penalty of byte-granularity heap randomization. The design of RUMA is based on per-instruction micro-benchmarks of Intel ISA. 

To measure the performance impact of RUMA, we apply various allocators (RUMA, tcmalloc, dlmalloc, jemalloc and ottomalloc) to SPEC2006 and compare their benchmark results. To test compatibility and analyze various correctness issues of byte-granularity heap randomization (which results misaligned memory access), we use Coreutils utilities test suite, Nginx web server test suite, and ChakraCore JavaScript engine test suite. We discuss the various issues regarding compatibility in~\autoref{s:compat} and summarize limitations in~\autoref{s:limit}.

\begin{figure*}[!t]
	\centering
	\includegraphics[width=0.9\textwidth]{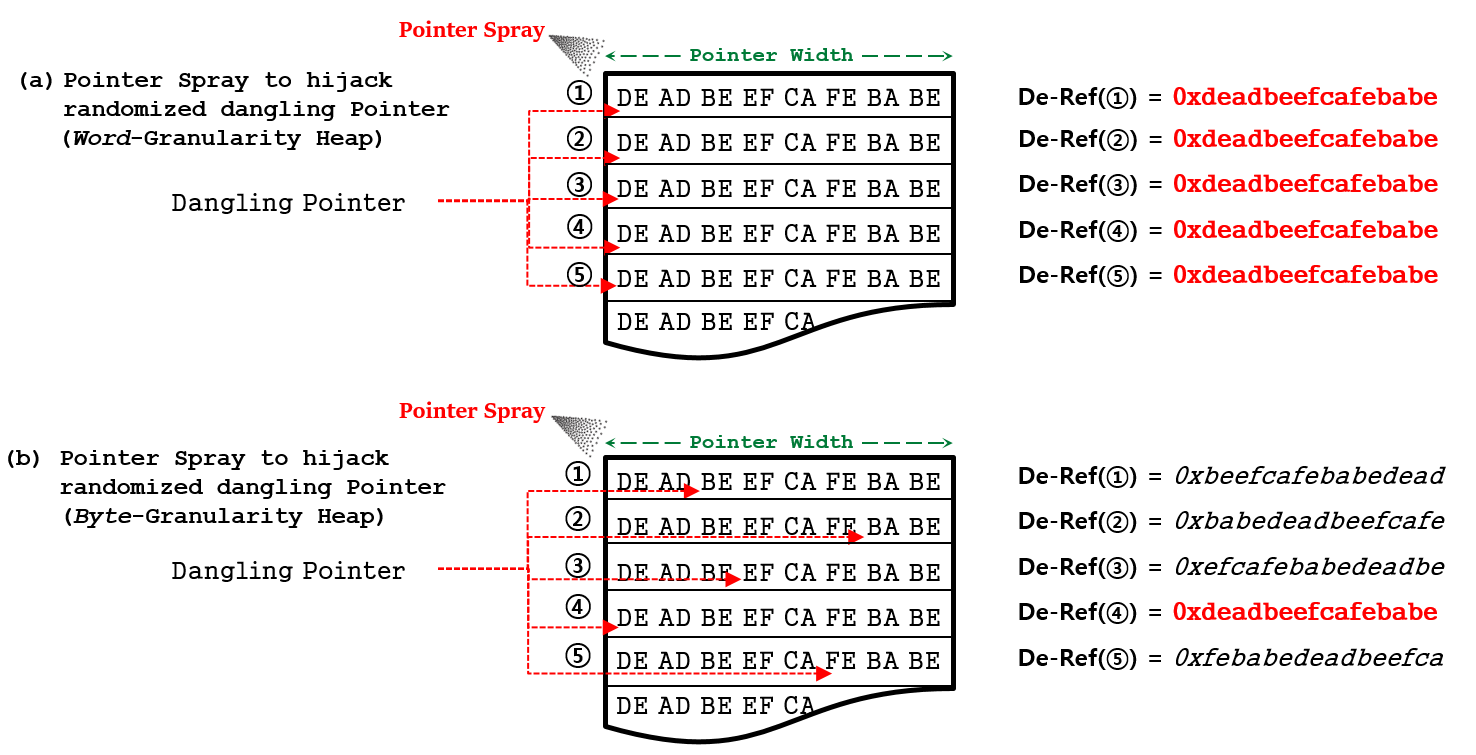}
	\caption{Simplified example cases of dereferencing a dangling pointer for Use-After-Free under pointer-spray attack (assuming \texttt{0xdeadbeefcafebabe} is the crafted pointer). The box in the middle represents dangling pointed object and each row indicates pointer-type member variable. Assume there are five possible dangling pointers due to randomization. For better visualization, the memory dump is shown in big-endian format.}
	\label{f:uafwithoutpar}
\end{figure*}

\section{Security Analysis of Byte-Granularity Heap Randomization}
\label{s:main}

In modern heap exploitation, achieving direct manipulation of code pointer is unlikely due to the state-of-the-art compiler defenses such as VTGuard, pointer encryption, and VT-read-only checks~cite{sarbinowski2016vtpin}. As a result, the surviving remnants of heap exploitation techniques often target data pointer and involve convoluted steps of pointer manipulation to complete the exploit. The key procedure is controlling the program to use the attacker-crafted non-pointer values as an intact pointer. Exploitation of typical heap vulnerabilities such as use-after-free, out-of-bounds access (including heap overflow), type confusion and uninitialized access mostly involves multiple steps of \emph{invalid pointer dereference} to trigger information leakage, achieve arbitrary memory access and finally, execute arbitrary code.

Traditional heap spray places a huge amount of NOP-sled and shellcodes into predictable heap address\footnote{Heap address prediction via massive heap spray is feasible under 32-bit address space layout randomization. In case of 64-bit address space layout randomization, heap spray becomes equally feasible if a heap segment base address is exposed.} as direct code pointer manipulation was easy, however, the goal of modern heap spray is more focused on placing crafted pointers around out-of-bounds heap area. Because heap spray allows an attacker to control a broad range of heap region, malicious pointer-spraying could be a threat without any (for 32-bit address space) or with only limited (for 64-bit address space) information disclosure. In this section, we first clarify the attack model and assumptions, then discuss the security effectiveness of byte-granularity heap randomization in three terms: (i) successful triggering of heap vulnerability (ii) information leakage attacks (iii) bypassing byte-granularity heap randomization.

\subsection{Attack Model and Assumption}
The goal of byte-granularity heap randomization is to hinder the \emph{pointer-spraying} which is often a critical methodology for building multiple-staged exploits. \emph{Pointer spraying} is an attack technique that constructs out-of-bounds access data (could be read by benign code or written against benign data) with a repeated sequence of pointers. Because the attacker can guarantee that any pointer (and 64bit variables) inside heap must follow specific address alignment (e.g., \texttt{sizeof(void*)}), spraying the same byte sequence of any 64bit variable allows correct dereferencing of out-of-bounds access if the spray region overlaps with the out-of-bounds access area. This allows an attacker to manipulate pointers without pinpointing its location or predicting the exact distance between dangling-pointer fake objects. 

Our discussion is based on the assumption that the attacker initially builds the exploit chain based on \emph{limited address layout prediction}. We assume memory disclosure capability that the attacker initially has provides limited information regarding address layout. For example, an information leakage bug can expose some pointer values that allows attacker to calculate corresponding segment base addresses (e.g., a code pointer reveals the base address of one code segment); and sophisticated heap allocation control based on object allocator analysis allows attacker to predict the \emph{relative} heap memory layout (chunk sequence, distribution, and so forth). The attack model does not assume that the attacker initially has the capability of thoroughly arbitrary/repeatable memory disclosure; which reveals the entire address space equally as debugging the application. Byte-granularity heap randomization augments the existing defense with finer-granularity, and hence we assume the presence of adequately configured and applied ASLR and Data Execution Prevention (DEP) as the previous study does. Finally, we expect that our attacker feeds untrusted input (e.g., PDF document, JavaScript, Network Stream) to the corresponding application parser that has heap vulnerabilities.

\subsection{Successful Triggering of Heap Vulnerabilities}
Any triggering step of heap vulnerabilities that occurs due to \emph{out-of-bounds} access\footnote{In this paper, out-of-bounds access indicates memory access that crosses heap chunk bound.} are affected by byte-granularity heap randomization. For example, the first use of dangling-pointer in use-after-free guarantees to crash any application with 87.5\% (75\% in 32-bit) probability as there are eight (four in 32-bit) possible outcomes of the misinterpreted pointer alignment.

Consider the exploitation steps of use-after-free: (i) an object is \emph{freed} and a dangling pointer is created, (ii) the attacker places a crafted object around the dangling-pointed memory region, and (iii) the program \emph{uses} the dangling pointer as if the original object member variables (pointer member variables) are still intact thus using attacker's crafted pointer. These steps imply that there are two independent heap chunk allocations around the dangling-pointed heap area. Although the address of each heap chunks is random, if the allocation granularity is bigger than the pointer-width, an attacker can spray the heap and overlap the fake object and dangling-pointer thus successfully trigger the use-after-free without pinpointing the exact memory addresses.

This effectiveness can be described by depicting a simplified example. \autoref{f:uafwithoutpar} depicts an example case of dereferencing a dangling pointer (to access a pointer member variable) after attacker launches a pointer-spray attack. For simplicity, let's assume attacker wants to hijack a pointer into \texttt{0xdeadbeefcafebabe} and there are five unpredictable cases of dangling pointers which will be randomly decided at runtime.

In \autoref{f:uafwithoutpar}a, an attacker can hijack the target pointer member variable with a very high chance because the heap randomization follows word-granularity. The attacker can spray the eight-byte sequence ``\texttt{DE AD BE EF CA FE BA BE}'' sufficiently long to defragment the heap region and bypass the randomization. However in \autoref{f:uafwithoutpar}b, the randomization is byte-granularity thus the attack fails with 87.5\% probability regardless of the spray; unless the pointer is composed with same bytes (we discuss this issue at the end of this section).

The effectiveness of byte-granularity heap randomization is not specific to particular heap vulnerabilities. We emphasize that \emph{any exploitation step which involves the use of crafted pointer upon out-of-bounds heap access is affected}. For example, exploitation of heap overflow, uninitialized heap access vulnerability also involves out-of-bounds heap access~\cite{cve:2015-2411, cve:2016-0191} thus affected by byte-granularity heap randomization. 

So far, the security effectiveness of byte-granularity heap randomization seems small, as one out of eight (or four) triggering attempts will succeed. However, this probability of single dereferencing is not the probability of a successful attack. Modern heap exploitation usually involves multiple combination and repetition of such bug triggering. According to Google Project-Zero, successful exploitation of CVE-2015-3077 required up to 31 times of pointer confusion. As heap exploitation involves multiple uses of crafted pointers, the defense probability will increase exponentially. However exact calculation of defense probability based on a number of the crafted pointer is unrealistic as modern heap exploitation usually achieves complete information leakage capability in the middle of the exploitation. In next subsection, we discuss the effectiveness of byte granularity heap randomization considering information leakages.

\subsection{Information Leakage}
Information leakage vulnerabilities are typically discovered in software that implements ECMAScript~\cite{ecma} parsers including ChakraCore, V8, and Spider Monkey. It is well established that \emph{complete disclosure of memory} renders all the randomness-based exploit defenses ineffective; thus we confined the attack model to which the attacker is initially incapable of such capability. However, information leakage attack can be divided into many types, and each of them involves various triggering of heap bugs. In this subsection, we discuss the impact of byte-granularity heap randomization considering limited information leakage attacks.

Fully repeatable/arbitrary information leakage primitive is achieved by the special case of literal-array based OOB (Out-Of-Bounds read/write) vulnerability. For example, if OOB has \emph{write} capability which can \emph{directly} manipulate a \emph{backstore} pointer\footnote{Backstore pointer is a leaf pointer for reading/writing memory. Taking control over the backstore pointer usually allows the attacker arbitrary memory read/write capability.} successfully; and if the attacker can repeat manipulating/using the manipulated backstore pointer, the attacker gets. Once the backstore pointer is fully controlled by the attacker, most defenses regarding memory corruption lose their meaning. 

\begin{figure*}[!]
	\centering
	\includegraphics[width=1.0\textwidth]{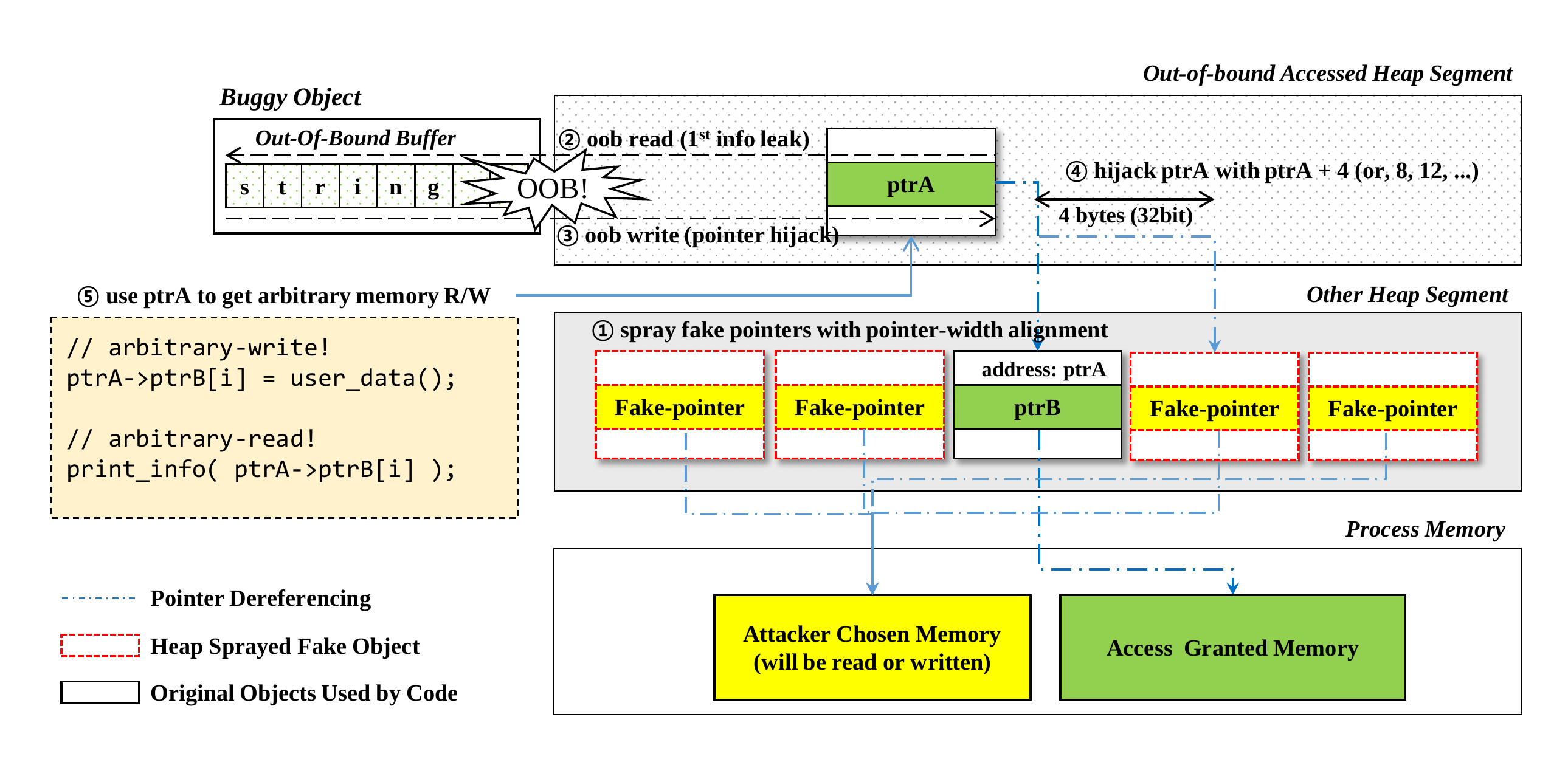}
\caption{Indirect (two-stage) manipulation of backstore pointer using OOB. If the heap segment that has backstore pointer (\texttt{ptrB}) and OOB accessed heap segment is different (which is likely), the attacker should spray crafted backstore pointers. This approach is highly dependent on the \emph{pointer-width} allocation granularity. Byte-granularity allocation breaks this approach with constant probability.}
	\label{f:leak}
\end{figure*}

However, taking control over the backstore pointer is not a straightforward task. Except for the lucky cases for the attacker, heap exploitation usually requires multiple steps of efforts to manipulate the backstore pointer. First of all, only a few objects have backstore pointer; therefore attacker should find such objects and control their allocation site and relative heap layout. Secondly, objects are isolated inside multiple heap segments depending on their types and sizes due to several heap isolation. Therefore, if an object with backstore pointer is not directly reachable via OOB, the attacker should target the parent pointer of such object.~\autoref{f:leak} depicts this with the simplified two-stage example.

\begin{figure}[!t]
	\centering
	\includegraphics[width=1.0\columnwidth]{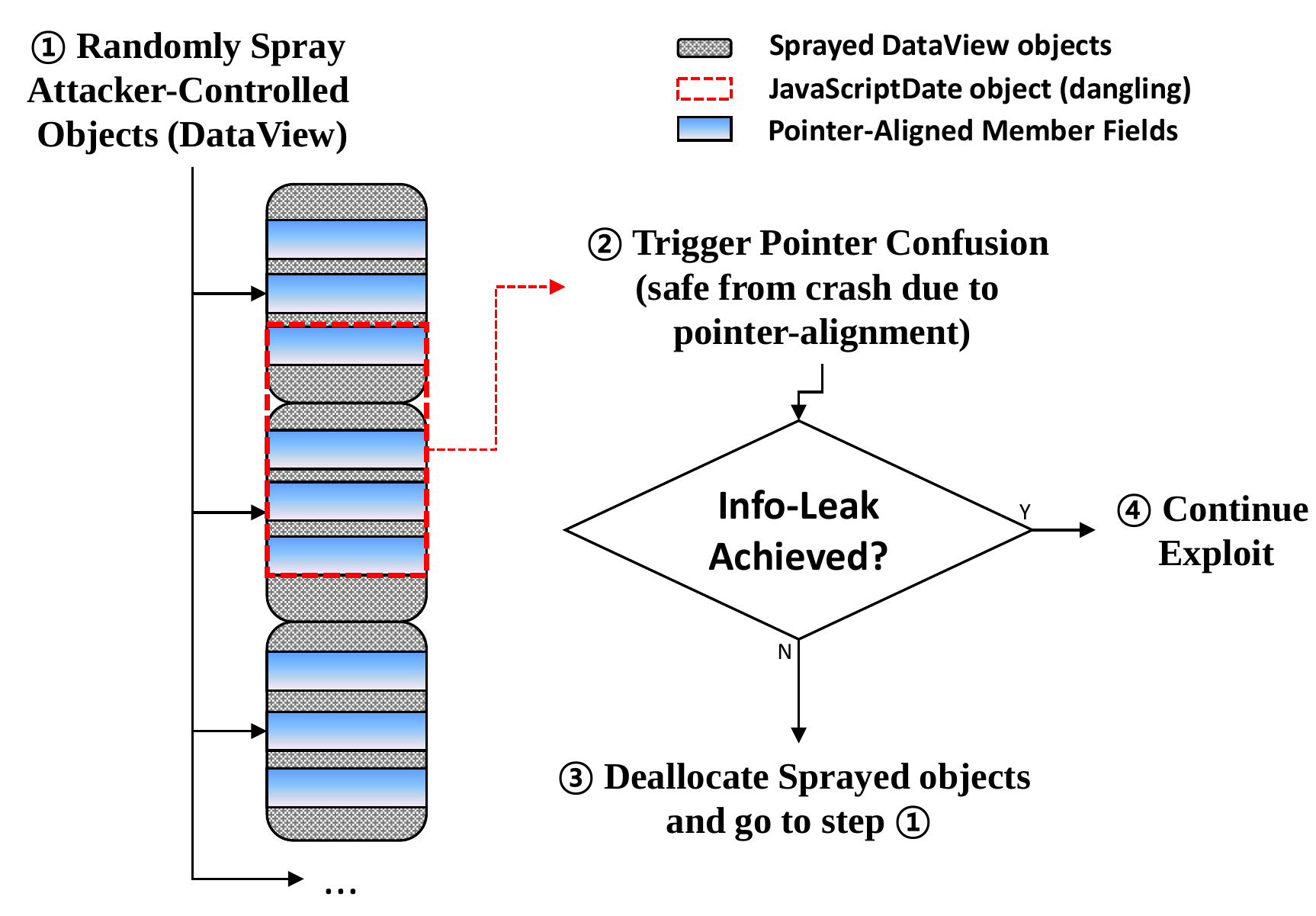}
	\caption{Information leakage steps of CVE-2016-0191.}
	\label{f:loki}
\end{figure}

Arbitrary controlled OOB, which we discussed so far, is, in fact, very ruthless exploitation primitive. In reality, not all OOBs can directly access the memory. For example, in case the OOB is based on \emph{object array}, an attacker cannot directly read/write the out-of-bounds memory contents. Since ECMAScript parsers do not provide any syntax that can directly dump the memory contents of an object (never should), additional work is required to turn such OOB into more useful exploitation primitive. What attacker can do here is to spray crafted objects around out-of-bounds object array, making the application to use a \emph{different type of object}. Using such object triggers further type confusions and enables an attacker to ultimately hijack a backstore pointer. This approach requires sophisticated object spraying and pointer-width heap allocation alignment.~\autoref{f:loki} effectively illustrates such an approach with the real-world example (CVE-2016-0191) where the crafted object spraying technique achieves the information leakage. The exploitation steps of \autoref{f:loki} were used in Pwn2Own 2016 for achieving information leakage against 64-bit Edge browser. In this exploitation, attacker achieves information leakage primitive by making the ChakraCore confuse the reference of the pointer inside \texttt{JavaScriptDate} object as the pointer of a \texttt{DataView} object. The exploitation \emph{repeats spraying} the crafted \texttt{DataView} object and using the dangling pointer of \texttt{JavaScriptDate} object until two object region overlaps with same pointer-width alignment.

\begin{table*}
	\centering
  	\footnotesize
	\begin{tabular}{|c|c|c|c|c|c|}
\hline
\# & CVE \#       					   & Bug Description  & Attack Target        & Remarks  \\ \hline \hline
1 & CVE-2013-2729 & out-of-bound     & Acrobat Pro (32bit)  & Pointer Spray  \\ \hline
2 & CVE-2015-2411 & use-before-init  & IE11 (32bit)         & Pointer Spray  \\ \hline
3 & CVE-2016-0191 & use-after-free   & Edge (64bit)         & Pointer Spray  \\ \hline
4 & CVE-2016-1653 & JIT compiler bug & Chrome (32bit)       & Pointer Spray  \\ \hline
5 & CVE-2016-1857 & use-after-free   & Safari (32bit)       & Pointer Spray  \\ \hline
6 & CVE-2016-5129 & out-of-bound     & Chrome (32bit)       & Pointer Spray  \\ \hline
7 & CVE-2012-4792 & use-after-free   & IE8 (32bit)          & Multiple Pointer Corruption  \\ \hline
8 & CVE-2013-0025 & use-after-free   & IE8 (32bit)          & Multiple Pointer Corruption  \\ \hline
9 & CVE-2014-3176 & out-of-bound     & Chrome (32bit)       & Multiple Pointer Corruption  \\ \hline
10 & CVE-2016-0175 & use-after-free   & Windows Kernel       & Multiple Pointer Corruption  \\ \hline
11 & CVE-2016-0196 & use-after-free   & Windows Kernel       & Multiple Pointer Corruption  \\ \hline
12 & CVE-2016-1017 & use-after-free   & Flash Player         & Multiple Pointer Corruption  \\ \hline
13 & CVE-2017-5030 & out-of-bound     & Chrome (64bit)       & Information Leak Analysis  \\ \hline
14 & CVE-2015-1234 & heap overflow    & Chrome (32bit)       & Non-Pointer Corruption  \\ \hline
15 & CVE-2017-2521 & out-of-bound     & Safari               & Non-Pointer Corruption  \\ \hline
16 & CVE-2016-1859 & use-after-free   & Safari               & Non-Pointer Corruption  \\ \hline
17 & CVE-2013-0912 & type confusion   & Chrome (32bit)       & In-bound Corruption  \\ \hline
18 & CVE-2017-0071 & type confusion   & Edge (64bit)         & In-bound Corruption  \\ \hline
19 & CVE-2016-1016 & use-after-free   & Flash Player         & None  \\ \hline
20 & CVE-2016-1796 & heap overflow    & OSX Kernel           & None  \\ \hline
\end{tabular}

	\caption{Summarized result of case study.}
	\label{t:cveresult}
	\vspace{-10pt}
\end{table*}

Finally, OOB is not the only vulnerability which results into information leakage capability. Less critical information leakage vulnerabilities exposes specific pointer values (attacker cannot choose these pointers) inside memory~\cite{cve:2016-1677,cve:2016-1665,cve:2016-1686} and do not provide arbitrary memory disclosure. With such limited information leakages, attacker can predict segment base addresses pointed by such pointers because ASLR is based on page-granularity. However, complete inspection of memory is not possible which is why attacker utilizes relative heap layout prediction and various spraying approaches in their exploit.

\subsection{Bypassing Byte-granularity Randomization}
Byte granularity heap randomization guarantees four (or eight) possible cases against any pointer manipulation due to out-of-bounds access. In turn, an attacker who wishes to hijack a single pointer (or any word-unit data), say, with a value of \texttt{0x1122334455667788}, stands a 87.5\% chance of failing in his/her overwrite attempt with pointer spraying (i.e., \texttt{0x8811223344556677}, \texttt{0x7788112233445566}, etc). Thus, a plausible way of bypassing byte granularity randomization is by constructing the entire chain of the exploit payload with \emph{byte-shift-independent values only}. A byte-shift-independent value is a word (or doubleword) composed of the same bytes (e.g., \texttt{0x9797979797979797} in a 64-bit system or \texttt{0x35353535} in a 32-bit system). 

At this point, the byte-shift-independent values are always invalid virtual addresses in current 64-bit system. In 32-bit system, such values can be predictable and valid address especially if the attacker allocates a sufficiently large region of the memory (i.e., allocating the lower 1-GByte of memory in a 32-bit address space will include an address of \texttt{0x35353535}). It could be a threat to byte-granularity heap randomization if attacker can place the crafted chunk at such address and construct the crafted pointers with byte-shift-independent values. Virtual addresses such as \texttt{0x35343534} make the same effect with 50\% probability. To address this issue, 32-bit version of RUMA provides configuration for avoiding such addresses similarly as EMET~\cite{dormann2014microsoft}. The difference between EMET and 32-bit RUMA is that while EMET initially pre-allocate (mmap) blacklist pages, 32-bit RUMA checks the address at allocation time and re-allocates the chunk. We compiled 32-bit version of SPEC2006 suites and tested performance impact of this algorithm. The algorithm caused up to 3\% overhead in 32-bit SPEC2006 benchmark. Detailed algorithm description for avoiding such addresses is described in~\autoref{s:peval}.

\begin{figure*}[!t]
	\centering
	\begin{subfigure}[t]{1.0\columnwidth}
    	\includegraphics[width=\linewidth]{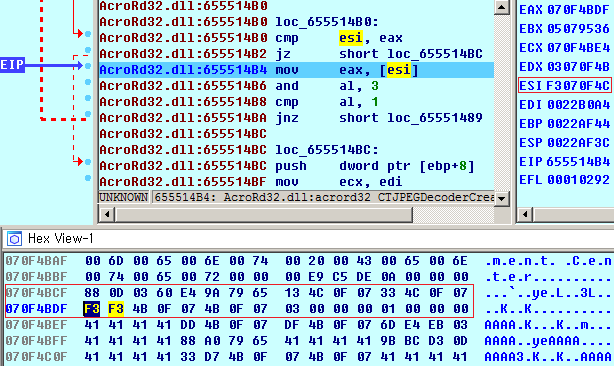}
            \caption{CVE-2013-2729}
    	\label{f:cve-2013-2729}	   
	\end{subfigure}
	\hfill 
	\begin{subfigure}[t]{1.0\columnwidth}
	    \includegraphics[width=\linewidth]{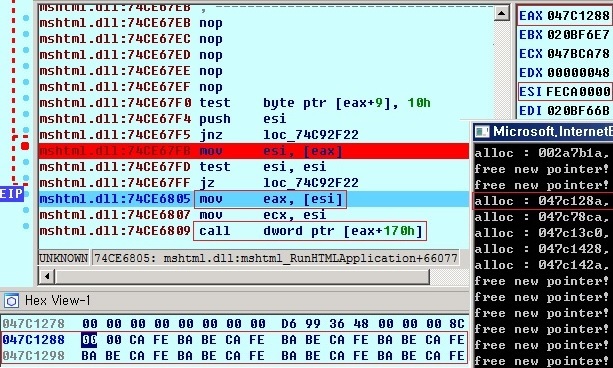}
             \caption{CVE-2013-0025}
	    \label{f:cve-2013-0025}
	\end{subfigure} 
	\caption{Reproducing CVE-2013-2729 (Acrobat Reader) and CVE-2013-0025 (Internet Explorer) exploitation for case study.}
	\label{f:cve}
\end{figure*}

\begin{figure*}[!t]
	\centering
	\begin{subfigure}[t]{1.0\columnwidth}
    	\includegraphics[width=\linewidth]{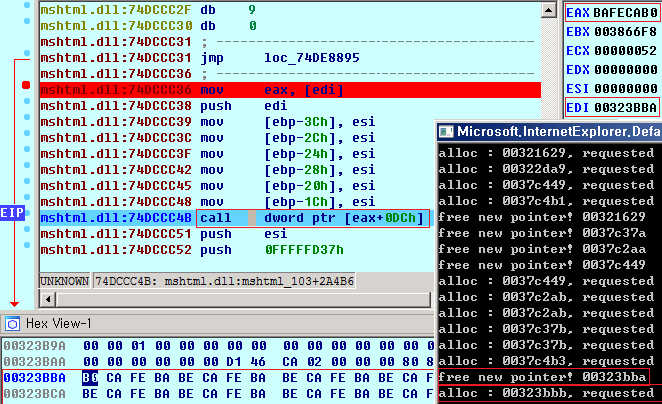}
            \caption{CVE-2012-4792}
    	\label{f:cve-2012-4792}	   
	\end{subfigure}
	\hfill 
	\begin{subfigure}[t]{1.0\columnwidth}
	    \includegraphics[width=\linewidth]{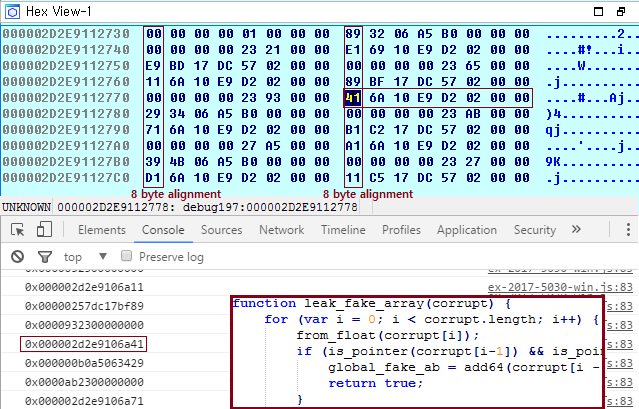}
             \caption{CVE-2017-5030}
	  \label{f:cve-2017-5030}
	\end{subfigure} 
	\caption{Reproducing CVE-2012-4792 (Internet Explorer) and CVE-2017-5030 (Chrome) exploitation for case study.}
	\label{f:cve2}
\end{figure*}

\section{Case Studies}
\label{s:seval}

Quantifying the probability of successful heap vulnerability exploitation is an arduous process and furthermore, proving the correctness of such probability (of successful exploitation) is practically impossible. Here we summarize case studies regarding the security effectiveness of byte-granularity heap randomization based on real-world exploitation analysis. For the analysis, we investigate publicly disclosed heap memory corruption vulnerabilities~\cite{cve:2016-1857,cve:2016-1859,cve:2016-1016,cve:2016-1017,cve:2016-1796,cve:2016-0175,cve:2012-4792,cve:2013-0025,cve:2015-2411,cve:2016-0191,cve:2015-1234,cve:2016-5129,cve:2016-1653,cve:2013-2729,cve:2014-3176,cve:2017-5030,cve:2017-0071,cve:2017-2521,cve:2016-0196,cve:2013-0912} that enabled attackers to achieve information leakage or arbitrary-code-execution against various software mostly from Pwn2Own contest and Google bug bounty program.

Throughout case studies, four of the attacks which we will discuss in detail~\cite{cve:2012-4792,cve:2013-0025,cve:2013-2729,cve:2017-5030} were fully reproduced and byte-granularity heap randomization was partially applied\footnote{Adopting byte-granularity heap randomization to legacy binary is not fully applicable except old version of Acrobat Reader due to compatibility issues which will be discussed later. In such case, we applied byte-granularity randomization only against exploit-relevant objects by identifying the object based on its size and allocation site for analysis. Details of such issues will be discussed in~\autoref{s:peval}}. Rest of the case studies were conducted based on debugging proof-of-concept attack codes with documented information. 

\subsection{Throwing off Pointer Spraying}
Throughout cases studies, we observed a number of \emph{pointer spraying} attempts for the successful attack. CVE-2016-0191 exploit extensively repeated spraying the crafted \texttt{DataView} object to trigger use-after-free. CVE-2016-1857 exploit sprayed crafted pointers to trigger use-after-free against \texttt{ArrayProtoType} object and hijack function call for \texttt{toString()}. CVE-2016-5129 sprayed crafted pointers around the out-of-bounds read heap region from the initial exploit step. CVE-2015-2411 exploit sprayed fake pointers around the uninitialized heap chunk of \texttt{Tree::TextBlock} object to initially trigger the vulnerability. CVE-2016-1653 exploit sprayed extensive amount of \texttt{Function} objects and hijacked the pointer inside an object using pointer-width allocation granularity.

\textbf{CVE-2013-2729} sprays a chunk of size 4096 bytes with fake pointers in it. An integer overflow vulnerability occurs while parsing a bitmap image embedded in a PDF document, which eventually leads to out-of-bounds heap access. As the full exploit code is available on the Internet, we reproduced the entire attack under debugging environment to see how byte granularity heap randomization hinders the exploitation. The PoC reliably achieves arbitrary code execution against 32-bit Acrobat Reader X in Windows 7; however, after we adopt byte-granularity randomness for heap allocation\footnote{by hooking the heap allocation APIs in the AcroRd32.dll import address table}, the PoC exploit constantly crashes due to dereferencing the invalid pointer (incorrectly aligned due to byte granularity randomization). However, the program never crashed while processing benign PDF documents. For example, the debug screen capture~\autoref{f:cve}a shows an example crash due to out-of-bounds heap pointer access (ESI is holding an invalid crafted pointer). The value of ESI is \texttt{0xF3070F4C}, which is a 1-byte shifted value from the intended one (\texttt{0x070F4C33}). This occurs because of the unpredictable change in the pointer-width alignment between the out-of-bounds buffer and the target heap chunk. Although this single step of pointer hijacking could be successfully operated in 25\% of the cases, overall attack chance is less than 1\% due to multiple needs of dereferencing out-of-bounds crafted pointers sprayed inside heap.

\subsection{Throwing off Multiple Pointer Corruption}
Some of the case studies often demonstrated that exploits require \emph{multiple pointer corruption}. CVE-2016-1017 exploit required two consecutive triggering of use-after-free pointer corruption. CVE-2016-0196 required three successive pointer corruption of use-after-free for the exploit. In the case of CVE-2012-4792 and CVE-2013-0025, the attacker must construct a multiple stage chain of crafted pointer dereferencing to ultimately change the virtual function call. CVE-2014-3176 exploit involves two consecutive pointer corruption of \texttt{Array} object and \texttt{ArrayBuffer} object to trigger initial memory read/write capability. CVE-2016-0175 uses two consecutive triggering of use-after-free. First use-after-free results arbitrary \texttt{or} operation with a constant number \texttt{2}, this operation triggers another use-after-free for secondary heap overflow vulnerability.

\textbf{CVE-2012-4792} use-after-free required chaining two crafted pointers. The vulnerability is caused by reusing a dangling pointer of \texttt{CButton} object after it has been freed.  The button object referenced by a dangling pointer in CVE-2012-4792 has a size of \texttt{0x86} bytes, and the use-after-free logic de-references this dangling pointer to retrieve VPTR inside the object. The exploit should manipulate two consecutive steps of pointer dereferencing: (i) manipulate V-Table pointer inside \texttt{CButton} object, (ii) manipulate function pointer inside V-Table. PoC code fails the first V-Table pointer manipulation (into test value \texttt{0xcafebabe}) with 75\% probability regardless of pointer spray. \autoref{f:cve-2012-4792} shows the debugging screen and memory allocation log trace of Internet Explorer while the PoC of CVE-2012-4792 is being triggered. From the figure, the allocation request for the attacker's data yields memory address \texttt{0x00323bbb} (3-byte distance from pointer-width alignment), which has a different alignment to that of the dangling pointer \texttt{0x00323bba} (2-byte distance from pointer-width alignment). Because of the discrepancy among these two memory alignments, the manipulated pointer via spraying becomes \texttt{0xbafecab0} which is an unintended pointer.

\textbf{CVE-2013-0025} is a use-after-free which required chaining three crafted pointers. Use-after-free in this case requires three successive manipulation of \texttt{CParaElement} pointer, V-Table pointer of \texttt{CParaElement}, and function pointer inside the V-Table. From the debugging screen of \autoref{f:cve}b, there are three successive pointer dereferencing before EIP is changed (e.g., \texttt{call dword ptr [eax+170h]}). The PoC exploit failed the first step of pointer hijacking ``\texttt{mov esi, [eax]},'' which was supposedly aimed at hijacking the pointer of \texttt{CParaElement} into a test value \texttt{0xcafebabe}, which is a sprayed pointer value.  At the bottom right of the debugging screen of \autoref{f:cve}b, there is an allocation log that reports that a heap chunk was yielded at \texttt{0x047c128a} (2-bytes off from word alignment). However, the value of the dangling pointer is \texttt{0x047c1288} (correct word alignment). 

\subsection{Throwing off Information Leak Analysis}
Byte-granularity heap randomization hinders the automation of \emph{information leak analysis}. Although the attacker can expose the memory contents via info-leak vulnerability, the attacker needs to interpret the semantics of leaked values. For example, if an attacker observes a value such as \texttt{0x12345678}, there should be an interpretation for this number (pointer, length, etc.). Pointer-width allocation granularity significantly helps this interpretation.

\textbf{CVE-2017-5030} is an OOB vulnerability which gives information leakage capability to the attacker. In this vulnerability, an attacker can print out the out-of-bounds accessed heap memory contents directly as float type numbers. Using this information leakage, the attacker gets useful information which will be later used. From the exploit codes, we can see that analyzing the leaked memory contents (\texttt{is\_pointer()}) in this attack leverages that heap chunks are always pointer-width aligned. Information leak analysis routine considers the leaked memory contents as 8-byte aligned array and examines the least significant bit of each element. If the least significant bit is set, attacker interprets the number as a pointer (V8 stores pointer in this way).~\autoref{f:cve2}b Shows memory dump of V8 and the leaked memory contents from Chrome debugging console while the exploitation is in progress. From the figure, the attacker assumes the leaked value \texttt{0x000002d2e9106a41} (highlighted with a box) is a pointer; based on the observation that the least significant byte among the 8-byte is an odd number. Although this vulnerability could still be exploited under byte-granularity randomized heap, automation of exploit becomes difficult.

\subsection{Non-Pointer and In-Bound Corruption}
Case study shows that there are attacks based on \emph{non-pointer manipulation} or \emph{in-bound corruption}. In such cases, byte-granularity heap randomization does not provide additional effectiveness over existing course-grained heap randomization. CVE-2015-1234 exploit targets corrupting the size member of an object via race-condition. CVE-2016-1859 exploit corrupts the \texttt{length} member variable of \texttt{GraphicsContext} object. In case of CVE-2017-2521, a JavaScript function \texttt{setPublicLength} fails to check security conditions thus allows increasing the length of an object. In-bound corruption is a memory corruption vulnerability that does not cross heap chunk boundary. For example, an object can have a buffer and other member variables together. In this case, the distance between the buffer and other member variable are fixed, regardless of heap randomization. Type confusion is another example of inbound corruption. CVE-2013-0912 and CVE-2017-0071 (type confusion vulnerability) results repeatable arbitrary memory read/write primitive using in-bound pointer corruption thus unaffected by heap randomization granularity.

Rest of the case studies were less affected by the heap randomization granularity because the vulnerability quickly achieves complete memory disclosure or the exploitation regarding pointer corruption is less complicated.~\autoref{f:cve} summarizes overall case study results.

\begin{figure*}[!t]
    \centering
    \includegraphics[width=\textwidth]{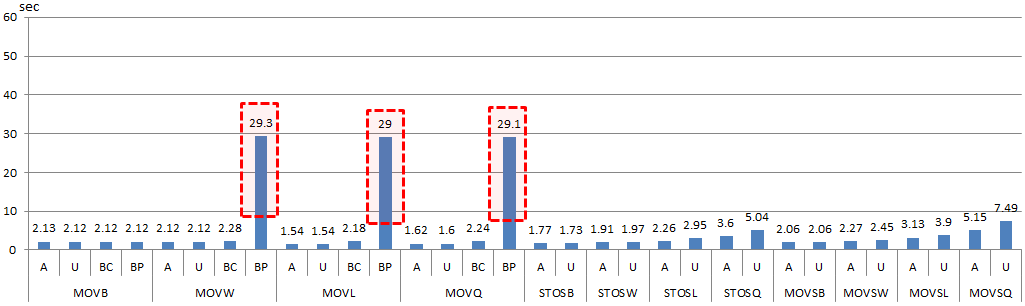}
    \caption{Instruction-level microbenchmark result for Intel i7-6500 (SkyLake). The Y-axis represents the time (consumed seconds) for repeatedly executing the unrolled instructions 134,217,728 (0x8000000) times. \texttt{A} stands for Aligned access, \texttt{U} stands for Unaligned access inside cache line, \texttt{BC} stands for unaligned access on Border of (L1) Cache line, \texttt{BP} stands for unaligned access on Border of Pages. We ran the same benchmark for additional 9 Intel CPUs (Core-i5, Xeon, and so forth) and verified similar results in all cases.}
    \label{f:microbench}
\end{figure*}

\section{Performance Analysis of Byte-Granularity Heap Randomization}
\label{s:peval}
As byte-granularity heap randomization inevitably involves CPU-level unaligned or misaligned memory access, we start this section by discussing some backgrounds regarding the unaligned access and analyze our findings in the performance and compatibility such as the atomicity and the alignment fault issues involved in the misaligned access. Based on microbenchmark analysis, we design a new memory allocator RUMA for efficient byte-granularity randomization. We further show in-depth testing results on RUMA using various benchmarks.

\subsection{Unaligned Memory Access}
The term \emph{alignment} can be used with many units such as page alignment, object alignment, and word alignment.  In this paper, the term alignment specifically refers to the \emph{CPU word granularity alignment}. Unaligned memory access can be observed in special cases in an Intel-based system (e.g., \texttt{\#pragma pack(1)} in the C language, x86 GLIBC I/O buffer handling). However, memory accesses are always encouraged to be aligned at multiples of the CPU word size. The main reason stems from the limited addressing granularity of the hardware. In general, CPU architectures feature a memory bus line with 8-byte addressing granularity; therefore, the retrieval of memory at an unaligned address requires exceptional handling from the hardware viewpoint. Handling such unaligned memory access may involve various performance penalty issues such as a possible delay of store-load forwarding, cache bank dependency, and cache miss. The major penalty induced by unaligned memory access is closely related to the increased number of L1 cache misses. Because the CPU fetches memory contents from a cache with \emph{cache line granularity}, unaligned memory access that crosses the boundary of two separate cache lines causes both cache lines to be accessed to retrieve a single word, causing a performance penalty.

Other than performance, unaligned memory access also raises concerns regarding memory access atomicity. Atomic memory access is a critical issue in concurrent software development. Multithreaded applications utilize various types of synchronization primitives such as mutex and semaphore. The key primitive of critical section (protected by the lock) is that the execution should be atomic from the perspective of each thread. Unaligned access raises concerns when the application uses thread synchronization, lock-free algorithms, or lock-free data structures, which relies on instruction-level atomicity such as the \texttt{InterlockedCompareExchange()} function. Because single unaligned memory access can split into multiple accesses at the hardware level, the atomicity of memory access may not be guaranteed. 

In fact, ARM architecture does not support such atomicity for unaligned access. Even recently, compare-and-swap (CAS) instructions in ARM (e.g., \texttt{ldrex, strex}) fail to operate if the target memory operand is unaligned. This is the major reason we conclude byte-granularity heap randomization is yet infeasible in ARM architecture. However, Intel microarchitecture (since P6) supports atomicity for such instructions even if the memory address is arbitrarily misaligned~\cite{intelmanualP6}. For example, CAS instructions of Intel ISA (e.g., \texttt{lock cmpxchg}) maintains memory access atomicity in all cases. The Intel official manual states that atomicity of \texttt{lock} is supported to arbitrarily misaligned memory. 

\begin{table}
    \centering
      \footnotesize
    \begin{tabular}{|c|c|c|}
\hline
OS & CR0.AM & EFLAGS.AC \\ \hline \hline
Windows & disabled & disabled \\ \hline
Linux & enabled & disabled \\ \hline
OSX & disabled & disabled \\ \hline
\end{tabular}

    \caption{Default configuration of CR0.AM bit and EFLAGS.AC bit in major operating systems. Unless two bits are enabled at the same time, unaligned access does not raise an alignment fault under the Intel architecture.}
    \label{t:trap}
    \vspace{-10pt}
\end{table}

Another important issue regarding unaligned memory access is the \emph{alignment fault}.  An \emph{alignment fault} (which raises SIGBUS signal in Linux) occurs in the event of unaligned data memory access, depending on the CPU configuration. There are two configuration flags regarding the alignment fault in the Intel architecture, namely the \texttt{AM} bit in the system \texttt{CR0} register and the \texttt{AC} bit in the \texttt{EFLAGS} register. Unless such bits are enabled together, Intel architecture does not raise an alignment fault. \autoref{t:trap} summarizes the default configuration of these registers in well-known operating systems.

\begin{figure*}[!t]
    \centering
      \footnotesize
    \includegraphics[width=\textwidth]{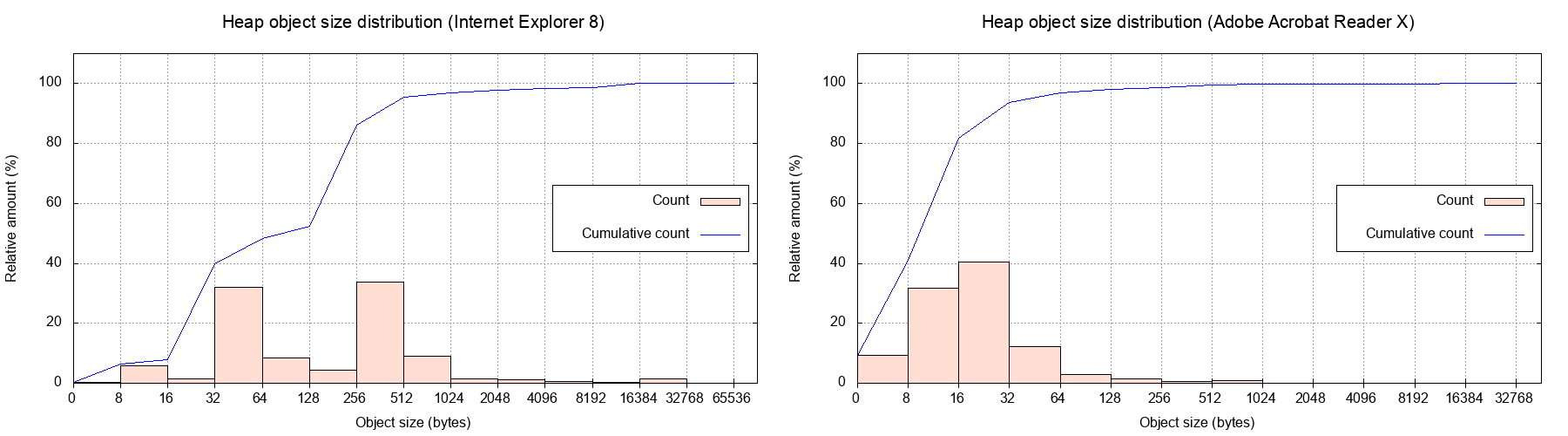}
    \caption{Size distribution of allocated heap objects in Internet Explorer 8 and Adobe Acrobat Reader X.}
    \label{f:heapdis}
    \vspace{-1pt}
\end{figure*}

\begin{table}
    \centering
      \footnotesize
    \begin{tabular}{|c|c|c|}
\hline
Configuration            & \# of Page-Faults & \# of Cache-Miss \\ \hline \hline
Fully Aligned            & 45               & 7,647 \\ \hline
Breaking Cache Line      & 46               & 7,527 \\ \hline
Breaking Page Border     & 45               & 76,181 \\ \hline
\end{tabular}

    \caption{PERF benchmark revealed the reason of exceptionally high performance penalty of unaligned memory access that crosses the border of two pages, which is the increased cache miss.}
    \label{t:perf}
    \vspace{-10pt}
\end{table}

However, there is an exception that raises an alignment fault regardless of such configuration. In the case of the Intel SSE~\cite{wiki:sse} instruction set, an instruction such as \texttt{MOVNTQDA} raises an alignment fault if the target memory operand address is misaligned at a \emph{16-byte boundary}. Normally, an arbitrary heap memory address has no guarantee to be 16 byte aligned in general. Therefore, codes that use SSE instruction do not assume that the given memory address for SSE operand will be 16-byte aligned. The pre-processing codes of SSE instruction checks the address alignment and properly handles the 128-bit misaligned memory portion. However, some of the compiler optimization cases (e.g., Clang -O3) predicts the chunk alignment and raises problem. We discuss more details regarding this issue with the compatibility experiment analysis.

\begin{table*}[!t]
	\centering
  	\footnotesize
	\begin{tabular}{|c|c|c|c|c|c|}
\hline 
CPU 			& OS 						& Compiler and Library Environment				 					& Penalty 	& Alignment Handling & IfAA/IfUA   \\ \hline \hline
Intel i7-6700  & Windows 10 				& VS2015 vcruntime140.dll (32bit)		& 5\%		& yes 	& REP/REP \\ \hline
Intel i7-6700  & Windows 10 			 	& VS2015 (64bit)						& 15\%		& yes 	& SSE/SSE \\ \hline
Intel i7-4980 	& Windows 8.1  		 		& VS2010 msvcr100d.dll (32bit) 			& -20\%		& yes 	& SSE/REP \\ \hline
Intel i7-4980 	& Windows 8.1  				& VS2010 msvcr100d.dll (64bit) 			& 0\% 		& yes 	& SSE/SSE \\ \hline
Intel i7-4980 	& Ubuntu 15.10 	 			& GCC 5.2.1 glibc-2.21 (32bit)			& 0\% 		& yes 	& SSE/SSE \\ \hline
Intel i7-4980 	& Ubuntu 15.10 	 			& GCC 5.2.1 glibc-2.21 (64bit)			& 0\% 		& yes 	& REP/REP \\ \hline
Intel i7-4980 	& Ubuntu 14.04  			& GCC 4.8.4 glibc2.19 (32bit) 			& 10\% 		& yes 	& REP/SSE \\ \hline
Intel i7-4980 	& Ubuntu 14.04 		 		& GCC 4.8.4 glibc2.19 (64 bit) 			& 5\% 		& yes 	& SSE/SSE \\ \hline
Intel i7-4980 	& OSX El Capitan 			& Apple LLVM 7.3 (32bit) 				& 0\% 		& yes 	& SSE/REP \\ \hline
Intel i7-4980 	& OSX El Capitan  			& Apple LLVM 7.3 (64 bit) 				& 5\% 		& yes 	& REP/REP \\ \hline
Intel i7-3770 	& Ubuntu 16.04 				& GCC 5.4.0 glibc-2.23 (64bit) 			& 0\% 		& yes 	& REP/REP \\ \hline		
Intel i5-3570 	& Fedora20 	 				& GCC 4.8.2 glibc-2.18 (64bit)			& 0\% 		& yes 	& SSE/SSE \\ \hline
Intel i5-3570 	& Debian7.5   				& GCC 4.7.2 glibc-2.13 (64bit)			& 0\% 		& yes 	& SSE/SSE \\ \hline
Intel i5-3570 	& FreeBSD9.1 (32bit)		& GCC 4.2.1 bsdlibc (32bit)				& 100\% 	& no  	& REP/REP \\ \hline
Intel i5-760 	& Ubuntu 12.04 Server  	  	& GCC 4.6.3 (32bit) 					& 0\% 		& yes 	& SSE/SSE \\ \hline
Intel i5-760 	& Ubuntu 12.04 Server  	  	& GCC 4.6.3 (64bit) 					& 0\% 		& yes 	& SSE/SSE \\ \hline
Intel i5-760 	& Windows 7   				& VS2010 msvcr100d.dll (32bit) 			& 50\% 		& yes & SSE/REP \\ \hline
Intel i5-760 	& Windows 7   				& VS2010 msvcr100d.dll (64bit) 			& 0\% 		& yes   & SSE/SSE \\ \hline
\end{tabular}

	\caption{Unaligned access penalty (in running time) of various memcpy implementations for 1-Mbyte buffer.  ``IfAA'' stands for: Instruction used for Aligned Address. ``IfUA'' stands for: Instruction used for Unaligned Address.}
	\label{t:memcpy}
	\vspace{-10pt}
\end{table*}

\subsection{Microbenchmark for Unaligned Access}
Unaligned access is often observed under Intel-based system. In recent Intel microarchitectures, the performance penalty of unaligned access is being reduced since Nehalem~\cite{nehalem}. Here, we show microbenchmark results regarding unaligned access in recent Intel CPUs. All the per-instruction benchmarks are composed of assembly code only, thus avoiding any compiler dependency.  We used ten different Intel x86-64 CPUs and measured the execution time of 134,217,728 (0x8000000) iterations for memory access instructions.

Performance measurement for instruction is dependent on CPU pipeline state. To make the worst-case CPU pipeline, we used the same instruction for 48 consecutive times and repeated such 48 consecutive execution using a loop. The loop was composed of \texttt{dec ecx} and \texttt{jnz}, both of which have lower instruction latency and reciprocal throughput compared to the memory access instructions. This configuration makes the worst-case CPU pipeline for memory access. \autoref{f:microbench} shows the overall results. In the benchmark, only the register and memory were used as target operands (the immediates were not used since the instruction latency was lower). 

Throughout the evaluation we have discovered that the performance penalty of unaligned access is severely biased by rare cases,~\autoref{f:microbench} is one of the experiment results. In particular, the performance penalty of unaligned access is 0\% if the accesses have entirely occurred inside the cache line. In case the unaligned access broke the L1 cache line, the performance penalty raised up to 30--70\%. In case the unaligned access broke the border of two 4-KByte pages, the performance penalty was suspiciously high (marked as red in the figure). In the case of REP-based instructions (REP counter value is 256), the performance penalty of unaligned access was mostly 20--30\%. To investigate the reason for the exceptionally high-performance penalty of unaligned access that crosses page border, we further used the PERF~\cite{perf} benchmark and found that the dominant factor is the increased cache miss.~\autoref{t:perf} summarizes the PERF benchmark conducted on the Intel i7-6700 Ubuntu14.04 Server in a 64-bit environment.

Thus far, the microbenchmark in ~\autoref{f:microbench}, suggests the performance penalty of unaligned access in modern Intel architecture is high only when the access crosses the border of two cache lines, and extremely high if the access crosses two-page boundaries.

\subsection{RUMA}

According to the cycle-level instruction benchmark, unaligned access in L1 cache line border and page border hindered the performance. To avoid such memory access while enabling byte-granularity heap randomization, we implemented a byte-granularity randomized heap allocator suited for Intel architecture, namely Randomly Unaligned Memory Allocator (RUMA). The goal of this allocator is to randomize the heap chunk location with \emph{byte granularity} while minimizing the unaligned memory access occurs at the border of L1 cache line and page boundary. We implemented RUMA as part of Clang runtime, and wrote an LLVM pass for automated allocator substitution.

To break the pointer-width allocation granularity, RUMA additionally allocates \texttt{sizeof(void*)} additional memory space in addition to original allocation request size to reserve \texttt{sizeof(void*)} dummy space. After the allocation algorithm selects the proper location for the new chunk, RUMA yields the address which is randomly increased in bytes between zero and \texttt{sizeof(void*)-1}. We not that the frontend implementation (adding random numbers) for applying byte-granularity randomization is a simple task and not our main contribution. The main contribution of RUMA is minimizing the impact of byte-granularity randomness based on the previously discussed analysis and experiments.

The backend allocation algorithm of RUMA is mainly based on \texttt{jemalloc}, where the heap space is organized as multiple pools each holding objects of certain size class. However, RUMA considers two special cases: (i) chunk size less than L1 cache line width, (ii) chunk size bigger than L1 cache line less than page size. The size of L1 cache line (usually 64-byte or 128-byte) is dynamically calculated during the allocator initialization and page size is statically assumed to be 4KB. Once the allocation size is determined, RUMA searches for an available chunk based on \texttt{jemalloc} algorithm with additional constraints that minimize the cases where chunks are spanning across special memory borders.

\begin{table*}[!t]
    \centering
    \footnotesize
    \begin{subtable}{.7\linewidth}
        \centering
        \begin{tabular}{|c|c|c|c|c|c|c|c|}
\hline
Benchmark  & Libc            &   RUMA           & ottomalloc      & jemalloc        & tcmalloc        & \# Allocs \\ \hline \hline
 perlbench &             264 &              263 &             330 &             259 &             252 & 358,141,317 \\ \hline
     bzip2 &             424 &              434 &             420 &             421 &             417 &         168 \\ \hline
       gcc &             242 &              238 &             318 &             231 &             233 &  28,458,531 \\ \hline
       mcf &             200 &              202 &             196 &             195 &             197 &           3 \\ \hline
     gobmk &             403 &              413 &             407 &             405 &             404 &     656,924 \\ \hline
     hmmer &             344 &              386 &             344 &             348 &             349 &   2,474,260 \\ \hline
     sjeng &             407 &              423 &             406 &             413 &             407 &           4 \\ \hline
libquantum &             354 &              356 &             352 &             353 &             353 &         179 \\ \hline
   h264ref &             440 &              529 &             441 &             440 &             447 &     177,764 \\ \hline
   omnetpp &             270 &              217 &             284 &             211 &             218 & 267,158,621 \\ \hline
     astar &             325 &              316 &             306 &             310 &             307 &   4,799,953 \\ \hline
 xalancbmk &             172 &              145 &             228 &             130 &             122 & 135,155,546 \\ \hline
\hline
Total (geomean) &    100.0\% &           99.7\% &         106.5\% &          94.8\% &         94.3\% &  \\ \hline
\end{tabular}

        \caption{Execution Time (sec)}
    \end{subtable}%
    \begin{subtable}{.3\linewidth}
        \centering
        \begin{tabular}{|c|c|c|}
\hline
Benchmark  & Libc & RUMA \\ \hline \hline
 perlbench & 648  & 631  \\ \hline
     bzip2 & 832  & 835  \\ \hline
       gcc & 865  & 868  \\ \hline
       mcf & 1638 & 1646 \\ \hline
     gobmk & 29   & 37   \\ \hline
     hmmer & 26   & 32   \\ \hline
     sjeng & 172  & 178  \\ \hline
libquantum & 95   & 105  \\ \hline
   h264ref & 64   & 80   \\ \hline
   omnetpp & 168  & 177  \\ \hline
     astar & 321  & 367  \\ \hline
 xalancbmk & 413  & 517  \\ \hline \hline
Total (geomean) & 100\% & 110.38\% \\ \hline
\end{tabular}
        \caption{Memory consumption (MB)}
    \end{subtable}
    \caption{SPEC2006 benchmark results. Various allocators (including RUMA) are applied to each program. While other allocators respect word-alignment for all chunks, RUMA randomizes the chunk location with byte-granularity.}
    \label{t:spec}
    \vspace{-10pt}
\end{table*}

\begin{table}
    \centering
      \footnotesize
    \begin{tabular}{|c|c|c|}
\hline
         & \multicolumn{2}{c|}{requests/sec} \\ \hline
         & Average & Standard deviation \\ \hline
Original & 31362.20   & 565.90             \\ \hline
Patched  & 31201.65   & 506.24             \\ \hline
\end{tabular}
    \caption{Performance impact of the Nginx patch.}
    \label{t:nginxbench}
    \vspace{-10pt}
\end{table}

For object allocation smaller than L1 cache line size (including the additional space for randomization), RUMA guarantees the memory location of the chunk to fit between two L1 cache line borders thus eliminate any performance penalty due to byte-granularity heap randomization. In case the requested size is bigger than L1 cache line and yet smaller than page, RUMA places the chunk between page boundaries therefore avoid page boundary access. If the allocation size is bigger than a page, there is no additional handling as baseline allocator guarantees minimal border access without any additional handling.

In case the application is 32-bit, RUMA uses address filtering algorithm to handle the case of byte-shift-independant-pointers~\autoref{s:main} which could bypass RUMA. In case the requested chunk size is larger than \texttt{0x01010101} bytes, it is impossible to remove the byte-shift-independent address from the virtual address mapping. However, in case the size is smaller than \texttt{0x01010101} bytes, 32-bit RUMA ensures there would be no byte-shift-independent address inside the allocated chunk. The address checking algorithm is executed after the chunk selection and right before delivering the chunk to the application. If the chunk includes 32-bit byte-shift-independent address (e.g., \texttt{0x11111111}), RUMA keeps the chunk internally and allocate a new chunk. The algorithm for address inspection is as follow: (i) Calculate most-significant-byte (MSB) of chunk start address. For example, if the chunk address is \texttt{0x12345678}, MSB is \texttt{0x12}. (ii) Check if MSB-only address (e.g., \texttt{0x12121212}) is in between chunk start and end. (iii) Increase MSB by one and repeat step (i), and (ii).

Overall, the efficacy of RUMA would be optimal when all objects are small (less than L1 cache-line would be ideal). In reality, however, there are large heap objects. As the size of an object becomes bigger, the chance of having a costly unaligned access will increase even though RUMA minimizes the occurrence of unaligned access. To investigate the chunk size distribution in common applications in general, we traced heap allocation and deallocation requests of Acrobat Reader and Internet Explorer. ~\autoref{f:heapdis} shows the object size distribution of live chunks when Acrobat Reader and Internet Explorer are running. The results are based on allocation/deallocation/reallocation call trace. The left side graph in the figure is the result of Internet Explorer after rendering the Google index page, and the right side graph is the result of Adobe Reader after rendering an ordinary PDF document.

~\autoref{f:heapdis} suggests that average heap chunk sizes are usually small. Byte granularity heap randomization using RUMA can fully avoid unaligned access penalty if all chunks are smaller than L1 cache line length. In case the chunk is larger than L1 cache line, it is inevitable to place a chunk across two cache lines. In general, large heap chunks are requested to allocate buffers that are often accessed by bulk-memory access APIs such as \texttt{memcpy}. To verify the detailed impact of byte-granularity heap randomization against memory intensive APIs and large buffers, we analyzed various versions of memcpy implementations and conducted experiments. In particular, we ran 100,000 iterations of 1-Mbyte memory copy operation using memcpy and compare the execution time between two cases: (i) source and destination addresses of 1-Mbyte buffer are word-aligned, (ii) source and destination addresses of the buffer is not word-aligned (byte-aligned). This experiment is \emph{NOT} designed to measure the performance of unaligned access at the instruction level. Rather, the purpose of this experiment is to measure the performance impact and compatibility of unaligned access against bulk memory access APIs.~\autoref{t:memcpy} summarizes the results.

In~\autoref{t:memcpy}, the performance penalty is negligible in most cases.  However, a severe performance penalty is observed from the case of i5-3570 FreeBSD 9.1; and ironically, the case of i7-4980 Windows 8.1 shows negative performance penalty. The cause of such peculiar results can be explained by the fact that \texttt{memcpy} chooses a different version of implementation at runtime, depending on various parameters such as the address alignment, CPU features, size of the buffer, and so forth. Aside from the case of FreeBSD 9.1, all version of memcpy implementation detected the unaligned address and optimized the performance by changing the alignment to be aligned before beginning the actual memory access.

\subsection{SPEC2006 Benchmark}
To measure the performance impact and memory usage of RUMA under memory intensive environment, we use SPEC2006. The SPEC2006 benchmark suite did not suffer compatibility problem after we applied RUMA. However, measuring the performance impact of RUMA should be carefully conducted because existing software conventionally assume word granularity heap alignment, therefore some code could show unexpected behavior. In addition, the program might use multiple allocators or custom allocators thus render the experiment inaccurate. 

Before applying RUMA allocator, we analyzed the source code of each benchmark suite to verify if the application is suited for the experiment. In case of 400.perlbench, custom heap allocator (\texttt{Perl\_safemalloc}) was used depending on the build configuration parameters. However, we confirmed that under our build configuration, the benchmark used default glibc allocator (\texttt{malloc}). Similarly, 403.gcc had multiple custom allocators (\texttt{xmalloc} and \texttt{obstack}) but we checked that under our build-environment, glibc allocator was used (\texttt{obstack} internally used \texttt{xmalloc}, \texttt{xmalloc} internally used \texttt{glibc} allocator). Other benchmark suites had no particular issues. All benchmark suite programs were also dynamically analyzed to confirm how many heap chunks are affected.

To change the allocator of SPEC2006, we made an LLVM pass \textit{Ruma} and used \texttt{replaceAllUsesWith} LLVM API to replace glibc C/C++ allocators such as \texttt{malloc}, \texttt{free}, \texttt{\_Znwj}, and \texttt{\_ZdlPv}. To measure the performance impact, we applied RUMA as well as other open-source allocators to SPEC2006. The open-source allocators we used are \texttt{dlmalloc} (glibc), \texttt{tcmalloc}, \texttt{jemalloc}, and \texttt{ottomalloc}. The benchmark was conducted the under following environment: Intel(R) Xeon(R) E5-2630 CPU with 128GB RAM, Linux 4.4.0 x86-64 and Ubuntu 16.04. We used the glibc allocator as the baseline of performance.~\autoref{t:spec} summarizes the results. 

\begin{table*}[!t]
    \centering
      \footnotesize
    \begin{tabular}{|c|c|c|}
\hline
Application          &  \# failures       & Remarks           \\ \hline \hline
Coreutils            &  0/476             & 56 cases initially failed due to allocator substitution problem (fixed later) \\ \hline
Nginx      &  324/324           & All cases failed due to least significant bit (LSB) of pointer utilization issue \\ \hline
Nginx (patched) &  0/324             & Re-factorization of LSB pointer issue made Nginx fully compatible \\ \hline
ChakraCore           &  172/2,638         & Alignment Fault while using SSE (MOVAPS) \\ \hline
ChakraCore           &  176/2,638         & Assertion failure (alignment check) \\ \hline
ChakraCore           &  3/2,638           & Futex system call failure (unaligned parameter) \\ \hline
\end{tabular}
    \caption{Compatibility Analysis Summary.}
    \label{t:compat}
    \vspace{-10pt}
\end{table*}

\section{Compatibility Analysis of Byte-Granularity Heap Randomization}
\label{s:compat}
Modern software and system implementations often assume and require the heap allocation alignment to be word aligned. Breaking this assumption can affect existing software in various aspects. In order to adopt byte-granularity heap randomization as practical defense, implementation conflicts regarding heap alignment should be addressed. To analyze the compatibility issues of byte-granularity heap randomization, we conducted experiments with various applications/benchmarks.~\autoref{t:compat} summarizes the analysis results regarding various compatibility issues.

\subsection{Coreutils}
After substituting Coreutils \cite{coreutils} glibc allocator to RUMA, we ran the Coreutils test suite to see if there are compatibility issues. In our initial experiment, 56 out of 476 Coreutils test cases did not pass the test. It turned out that programs using the following APIs crashed during execution: \texttt{strdup}, \texttt{strndup}, \texttt{getline}, \texttt{getdelim}, \texttt{asprintf}, \texttt{vasprintf}, \texttt{realpath}, \texttt{getcwd}. After analyzing the root cause, we found that the reason was irrelevant to the byte-level allocation granularity. While our LLVM pass replaces the allocator calls in Coreutils, allocator calls inside libc remained. Since the above-mentioned libc APIs internally use default glibc allocator, two allocators has conflicted. To handle this issue, we ported such APIs to use RUMA allocator and extended the LLVM pass to replace such API calls as well. After porting the above-mentioned APIs for RUMA, all programs passed the test without causing any compatibility issues.

\subsection{Nginx}
We ran Nginx test suite after applying RUMA allocator. In our initial experiment, none of the 324 \texttt{nginx-tests} suite passed the check. According to our analysis, the root cause of the problem was the implicit assumption of heap chunk alignment in Nginx implementation. Because the code assumes that any heap object will be word-aligned, on some occasions the program chose to store boolean information inside the least significant bit (LSB) of pointers, perhaps for performance reason. That is, instead of declaring a boolean member in a structure or class, the boolean value is saved in the LSB of the object pointer.

\begin{lstlisting}[caption={Representative parts of the Nginx patch. LSB storage is replaced by a new member in \texttt{ngx\_connection\_s}.}, language=diff, style=custompy, label={l:diff}]
diff --git a/src/core/ngx_connection.h b/src/core/ngx_connection.h
index e4dfe58..5c15ca2 100644
--- a/src/core/ngx_connection.h
+++ b/src/core/ngx_connection.h
@@ -119,6 +119,7 @@ typedef enum {
 struct ngx_connection_s {
+    unsigned           instance:1;
     void               *data;
     ngx_event_t        *read;
     ngx_event_t        *write;
diff --git a/src/event/modules/ngx_epoll_module.c b/src/event/modules/ngx_epoll_module.c
index 76aee08..da948f2 100644
--- a/src/event/modules/ngx_epoll_module.c
+++ b/src/event/modules/ngx_epoll_module.c
@@ -618,7 +620,8 @@ ngx_epoll_add_event(ngx_event_t *ev, ngx_int_t event, ngx_uint_t flags)
 #endif

     ee.events = events | (uint32_t) flags;
-    ee.data.ptr = (void *) ((uintptr_t) c | ev->instance);
+    c->instance = ev->instance;
+    ee.data.ptr = c;

     ngx_log_debug3(NGX_LOG_DEBUG_EVENT, ev->log, 0,
                    "epoll add event: fd:%d op:%d ev:%08XD",
@@ -836,8 +841,7 @@ ngx_epoll_process_events(ngx_cycle_t *cycle, ngx_msec_t timer, ngx_uint_t flags)
     for (i = 0; i < events; i++) {
         c = event_list[i].data.ptr;

-        instance = (uintptr_t) c & 1;
-        c = (ngx_connection_t *) ((uintptr_t) c & (uintptr_t) ~1);
+        instance = c->instance;

         rev = c->read;
\end{lstlisting}

To handle this compatibility problem, we patched the Nginx-1.13.12 source code as shown in ~\autoref{l:diff}. After the patch, all test suite passed the check. To evaluate the performance impact of RUMA-compatible modification against Nginx, we benchmarked the request throughput using \texttt{wrk}~\cite{wrk}. The benchmark was repeated 10 times for each case.~\autoref{t:nginxbench} summarizes the benchmark result. According to the benchmark, no significant performance degradation was introduced by the modification.

\subsection{ChakraCore}
ChakraCore revealed interesting compatibility issues of byte-granularity heap randomization. We applied RUMA to ChakraCore then ran the standard test suite provided by the ChakraCore. The initial result indicated that all test cases failed to pass the check with the same error. The failure seemed irrelevant to the test case. We found that the initialization of ChakraCore JSRuntime accessed RUMA-affected chunk, however, the \texttt{-O3} optimization of Clang aggressively assumed the alignment of the heap chunk (assuming it would be 128-bit aligned) then used SSE instructions that require specific memory alignment such as \texttt{movaps} for fast execution. After changing the optimization level to \texttt{-O0}, 351 over 2,638 test cases (\texttt{interpreted variant}) failed to pass the check. 

Among the failures, 172 cases were caused due to SSE instruction alignment fault and 176 cases failed due to the assertion failure that explicitly requires word alignment for heap pointers before further operation. Interestingly, three cases failed due to time-out. After further analysis, we found that the failure was due to \texttt{futex} system call failure. Unlike \texttt{pthread mutex}, which is based on user-level \texttt{lock} prefixed instructions, the \texttt{futex} is based on Linux kernel system call which requires word-aligned address for its parameter. Because of this, the test suite failed to operate correctly as the system call raised an error. After finding this issue, we investigated the current (4.x) Linux kernel system calls to find a similar case. The \texttt{futex} was the only one that caused the problem with byte-granularity heap randomization. Overall compatibility analysis indicates that byte-granularity heap randomization requires high deployment cost.

\section{Limitation}
\label{s:limit}

\textbf{RUMA for ARM}. The performance impact of unaligned access is a serious issue for RISC processors such as ARM. In general, unaligned memory access is strongly discouraged in RISC architectures\footnote{PowerPC architecture can have 4,000\% penalty in the worst case of unaligned access~\cite{data-alignment}.}. In fact, ARM architecture started to support hardware-level unaligned access for some instructions (e.g., \texttt{LDR}) since ARMv6~\cite{armv6}. To investigate the feasibility of byte-granularity heap randomization in ARM, we conducted per-instruction memory access benchmark against Cortex-A9 and Cortex-A17.~\autoref{t:arm} summarizes the result.  According to our analysis, ARM architecture since ARMv6 indeed supports hardware level unaligned access. However, the support is for only two memory access instructions (\texttt{LDR, STR}) and high-performance penalty is observed at every 8-byte address border regardless of L1 cache or page size. Instructions such as \texttt{LDM} shows over 10,000\% performance penalty for unaligned access. The reason for such high penalty is due to the lack of hardware support. Since the hardware is incapable of executing \texttt{LDM} with unaligned memory operand, hardware raises fault signal and kernel emulates the instruction. In case of the \texttt{VLDR}, even emulation is not supported by the kernel. Therefore the execution fails on unaligned memory operand. Most importantly, unaligned memory operand does not support \texttt{LDREX} instruction which is required for instruction level atomicity. For such reasons, ARM based system is inappropriate to consider byte-granularity heap randomization at this point.

\textbf{Side Channel Attack}. With byte-granularity heap randomization, heap pointers do not follow word-granularity. In average, 75\% of heap pointers are not word-aligned. Assuming if the attacker is somehow able to measure the performance of dereferencing heap pointers precisely, she might be able to tell that some of them are misaligned around particular memory border. For example, the attacker can guess that a heap pointer is spanning across page boundaries while being unaligned if the access speed is relatively slow. So far, we fail to find any useful attack scenario by identifying such pointers. However, in theory, this can be considered as a potential side channel attack against byte-granularity heap randomization.

\textbf{Implementation Conflicts}. The adoption of byte-granularity heap randomization creates various implementation conflicts as discussed in~\autoref{s:compat}. One of the major cases among them is the use of LSB portion of heap pointer assuming the pointer is word-aligned. In~\autoref{s:compat}, we used Nginx for discussion; however we also found this issue in other applications as well. For example, Internet Explorer 11 uses the same implementation approach to mark the chunk type (Isolation Heap). Any programming techniques that rely on the assumption that \emph{the heap chunk has specific alignment} cannot be applied with byte-granularity heap randomization at the same time. In addition, \texttt{futex} is currently incompatible with RUMA as it require word-aligned address (other 4.x Linux system calls are not affected by alignment). Admittedly, the implementation compatibility issues are the major limitation for adopting byte-granularity heap randomization in practice as it requires significant engineering effort. However, we believe this is not a fundamental limitation that undermines the worth our research.

\begin{table*}[!t]
    \centering
      \footnotesize
    \begin{tabular}{|c|c|c|c|}
\hline
Architecture  &  Instruction                 & Penalty           & Remarks \\ \hline \hline
Cortex-A9     &  LDR/STR          & 100\%            & penalty occurs at 8-byte border \\ \hline
Cortex-A9     &  LDRB/STRB        & 0\%              & no penalty \\ \hline
Cortex-A9     &  LDM/STM          & over 40,000\%    & penalty always occurs, kernel emulation \\ \hline
Cortex-A9     &  LDM/STM (ThumbEE) & 7,000\%          & penalty always occurs, kernel emulation \\ \hline
Cortex-A9 (,A17)     &  VLDR/VSTR        & N/A              & alignment fault \\ \hline
Cortex-A9 (,A17)    &  LDREX/STREX      & N/A              & alignment fault (no atomicity) \\ \hline 
Cortex-A17    &  LDR              & 100\%            & penalty occurs on 8-byte border \\ \hline
Cortex-A17    &  STR              & 50\%             & penalty occurs on 8-byte border \\ \hline
Cortex-A17    &  LDM/STM          & over 2,000\%     & penalty always occurs, kernel emulation \\ \hline
\end{tabular}
    \caption{Per-instruction benchmark against ARM CPUs. The benchmark methodology is same to the Intel version.}
    \label{t:arm}
    \vspace{-1pt}
\end{table*}

\textbf{Information Disclosure using Byte-shift-independent Non-pointer Values}. Byte granularity heap randomization imposes difficulty of hijacking pointers by breaking the \texttt{sizeof(void*)} allocation granularity of randomized chunk allocation. As the result of byte granularity randomness, an attacker cannot leverage \emph{pointer spraying} technique for bypassing the randomized memory layout. The only option for reliable attack (other than information disclosure) is to rely on byte-shift-independent values, which make it hard (if not impossible with careful heap management) to craft valid pointers. But this is not the case for \emph{byte-shift-independent non-pointer values}, which can allow an attacker to craft reliable memory corruption as intended and then escalate the attack further. A representative example would be string length corruption~\cite{cve:2015-8651,cve:2013-0634} mentioned earlier in the paper~\autoref{s:main}.

\section{Related Work}
\label{s:relatedwork}

\textbf{HeapTaichi}.~\cite{ding2010heap} shows various heap spraying techniques that leverage the allocation granularity of memory allocators. For example, if allocation granularity is fixed, an attacker can split nop-sleds into several pieces and stitch them together with jump instructions. HeapTaichi claimed that reduced allocation granularity in heap randomization is better for security. However, the minimal allocation granularity considered in HeapTaichi is pointer-width. Although HeapTaichi discussed in-depth heap allocation granularity issues, no discussion regarding byte-granularity allocation and its ramification regarding security/performance was made.

\textbf{Address Space Layout Permutation}.~\cite{kil2006address} adopts a high degree of randomness compared to the existing ASLR and also performs fine-grained permutation against the stack, heap, and other memory mapped regions. Heap randomization is not the main theme of this work. However, the paper includes descriptions regarding fine-grained heap randomization. To adopt fine-grained address permutation for a heap, a random (but page-aligned) virtual address between \texttt{0} and \texttt{3} GB (assuming 32bit memory space) is selected for the start of the heap. Afterwards, a random value between \texttt{0} and \texttt{4} KB is added to this address to achieve sub-page randomization. According to this method, heap pointers should have random byte-level alignment, which involves unaligned access problem. However, discussion regarding unaligned access (due to byte-level randomization) or the security effectiveness of byte-granularity randomization was not discussed despite ASLP covered a broad range of fine-grained randomization issues. 

\textbf{Address Space Randomization}.~\cite{giuffrida2012enhanced} introduced fine-grained Address Space Randomization (ASR) for various OS-level components including heap object allocation. In this work, heap layout is effectively randomized by prepending random size padding for each object and permuting the sequence of allocated objects. The paper comprehensively explores various memory layout randomization strategies and propose various ideas regarding live re-randomization. They implement each randomization policies by patching the kernel and dynamic loader, or using binary code translation techniques. However, the security and performance impact regarding byte-level memory layout randomization is not the main interest of the paper. The main focus of the paper is comprehensive OS-level ASR and live re-randomization with a minimal performance penalty. 

\textbf{Data Structure Randomization}. Data Structure Layout Randomization (DSLR)~\cite{lin2009polymorphing} randomizes the heap object layout by inserting dummy members and permuting the sequence of each member variables inside an object at compilation time. The size of randomly inserted garbage member variable is multiple of \texttt{sizeof(void*)} thus respecting CPU alignment. The goal of DSLR is to diversify the kernel object layouts to hinder the kernel object manipulation attack performed by rootkits; in addition to thwarting the system fingerprinting and kernel object manipulation attack which relies on object layout information.

\textbf{Cling}. The isolation heap protection approach separates the heap into the independent area so that objects are allocated at different parts depending on their types. Indeed, these approaches can be observed in both academia and industry. Cling~\cite{akritidis2010cling} identifies the call site of a heap chunk allocation request by looking into the call stack. If the chunk allocation request originates from the same call site, Cling considers the type of heap chunk to be the same, which indicates that it is safe to reuse the same heap area for those chunk requests. If the type is assumed to be different from two allocation requests, Cling does not allow the heap area to be reused between those requests. In practice, the heap isolation methods can frequently be observed in various security-critical software such as Internet Explorer, Chrome, and Adobe Flash~\cite{heapiso}.

\textbf{Other Heap Randomization Approaches}. Incorporating randomization into heap allocation has been discussed in numerous previous works. Some approaches, such as those of Bhatkar et al. and Qin et al., respectively randomize the base address of the heap, as shown in~\cite{bhatkar2003address,qin2005rx}. Others randomize the size of the heap chunks, word-granularity location, allocation sequence and so forth~\cite{kharbutli2006comprehensively,iyer2010preventing,bhatkar2005efficient,berger2006diehard,novark2010dieharder,valasek2012windows,haller2016metalloc}. From all these heap fortifications works including our paper, the purpose in adopting the notion as well as the implementation differs from each other. The advancement from previous works is that we show how byte-granularity heap randomization mitigates crafted pointer spray, then design an allocator that optimizes performance cost of byte-granularity heap randomization. Instead of locating the heap chunk at memory location unpredictable by an attacker, byte granularity heap randomization aim to obstruct heap exploits which require pointer-width allocation granularity.

\section{Conclusion}
\label{s:conclusion}
In this paper, we proposed byte-granularity heap randomization and discussed its efficacy in various aspects. At first glance, breaking the randomization granularity from word to byte can be considered trivial. However, this seemingly insignificant change in granularity opened up a surprising number of research issues. One of a skeptical matter is that it hinders the word-aligned performance optimization at the CPU level. To overcome such problem, we introduce RUMA: an allocator optimized for byte-granularity heap randomization. The design of RUMA leverages recent advancement of CPU architectures for handling the misaligned access. To overcome the misalignment problem of byte-granularity, RUMA considers particular allocation sites regarding cache line. We implemented RUMA as part of Clang runtime for various evaluations. The performance cost of RUMA's randomization for SPEC2006 is around 5\% on average compared to other allocators without randomization.

{\footnotesize \bibliographystyle{acm}
\bibliography{bib/scholar,bib/url}}

\end{document}